\documentclass{article}
\usepackage{geometry}                
\usepackage{graphicx}
\usepackage{amssymb}
\usepackage{epstopdf}
\usepackage{array}
\usepackage{subfigure}
\usepackage{url}
\usepackage{hyperref}
\usepackage{xspace}
\usepackage{lscape}
\usepackage{isorot}

\usepackage{color}

\definecolor{gray}{rgb}{0.7,0.7,0.7}

\usepackage{supertabular}
\usepackage{hhline}
\newcommand\arraybslash{\let\\\@arraycr}
\makeatother
\begin{document}

\title{On Creativity of Elementary Cellular Automata}

\author{Andrew Adamatzky$^1$ and Andrew Wuensche$^{1,2}$ \\
$^1$ University of the West of England, Bristol, UK\\
$^2$ University of Sussex, Brighton, UK
}


\maketitle

\begin{abstract}

We map cell-state transition rules of elementary cellular automata
(ECA) onto the cognitive control versus schizotypy spectrum phase
space and interpret cellular automaton behaviour in terms of
creativity. To implement the mapping we draw analogies between a
degree of schizotypy and generative diversity of ECA rules, and
between cognitive control and robustness of ECA rules (expressed via
Derrida coefficient). We found that null and fixed point ECA rules lie
in the autistic domain and chaotic rules are 'schizophrenic'. There
are no highly articulated 'creative' ECA rules. Rules closest to
'creativity' domains are two-cycle rules exhibiting wave-like
patterns in the space-time evolution.
 
\end{abstract}

\section{Introduction: On Creativity}

Creativity is ubiquitous yet elusive concept. Everyone knows what it
means to be creative, e.g. to be successful in problem-solving and
generation of novel thoughts~\cite{gordon_1961}, but few can define
creativity rigorously. Substantial progress has been achieved in the
fields of computational and psychological creativity.  Thus, Kowaliv,
Dorin and Korb studied creativity of graph-pattern generation and
progressed towards outlining creativity as based on a probability of
pattern emergence~\cite{dorin_2009, kowaliw_2009}. In this sense, a
system is creative if it produces a pattern where the likelihood of
emergence is small. Wiggins formalises Boden's
concept~\cite{boden_1998} of exploratory creativity as exploration of
a conceptual space~\cite{wiggins}; thought, a question could be raised
--- is creativity in the complexity of conceptual space or the search
engine? Another computational approach to creativity is a generation
of novelty via conceptual blending~\cite{pereira_2007, li_2012}, and
use of analog machines in evolutionary creation of cross-domain
analogies~\cite{baydin_2012}.

\begin{figure}[!tbp]
\centering
\includegraphics[width=0.85\textwidth]{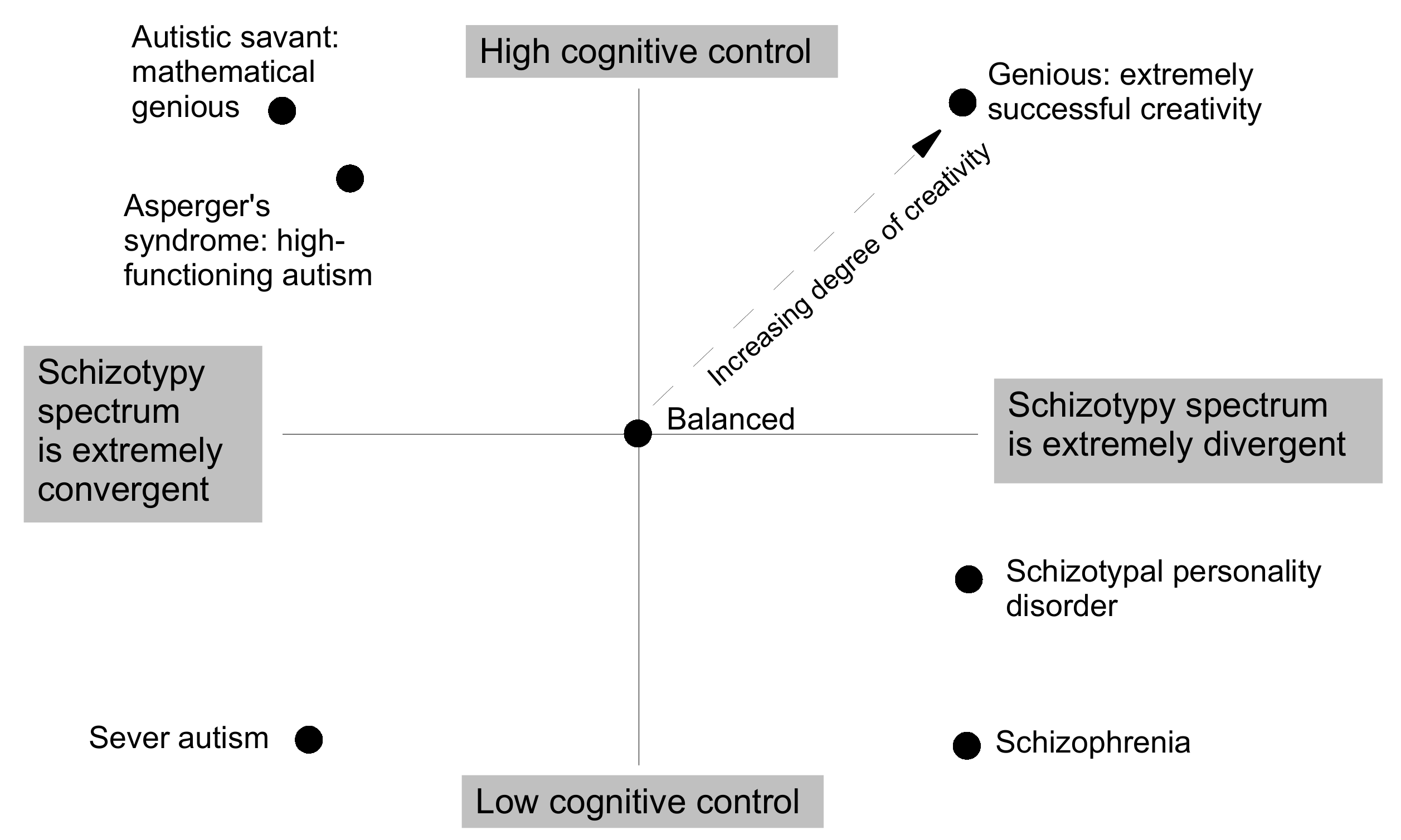}
\caption{Schyzotipy versus cognitive control spaces. Original scheme redrawn from Kuszewski's paper~\cite{kuszewski_2009}. }
\label{scheme}
\end{figure}

From a psychological and neurophysiological perspective there is a
great similarity between creativity and
psychoticism~\cite{eysenck_1992, eysenck_1993, abraham_2005, koh_2006}. The similarities include over-inclusive cognitive
style, conceptual expansion, associative thinking, and lateral
thinking dominating vertical (goal-oriented) thinking. In contrast to
creativity, however, psychoticism shows diminished
practicality~\cite{abraham_2005,koh_2006}.
Kuszewski~\cite{kuszewski_2009} provides plausible and psychologically
feasible indicators of creativity: divergent thinking and lack of
lateral inhibition; the ability to make remote associations between
ideas and concepts; the ability to switch back and forth between
conventional and unconventional ideations (flexibility in thinking);
generation of novel ideas appropriate for actualities; willingness to
take risks; and, functional non-conformity.  Cognitive control of
divergent thinking is a guarantee of creativity.  A person with
extremely divergent thinking yet unable to control will be a
`nutter'. Those who can fit their high schizotypy traits into a rigid
cognitive frame incline to genius.  Thus creativity could be
positioned together with autism and schizophrenia in the same `phase'
space (Fig~\ref{scheme}).

To develop cellular automata (CA) analogies of Kuszewsi's scheme we
assume that a cell neighbourhood configuration of a CA represents a
'thought', or some other elementary quantity of a mental process, and
a degree of schyzotipy is proportional to the diversity of global
configurations generated by the CA.  We can speculate that cognitive
control is equivalent to robustness of CA evolution. A cellular
automaton is robust if trajectory of a disturbed automaton, with some
cells' states changed externally, does not deviate, in terms of
Hamming distance, too far away from a trajectory of an undisturbed
automaton.  The degree of deviation caused by a disturbance is
measured by the Derrida coefficient.

\section{Elementary cellular automata}

An elementary cellular automaton (ECA) is a one-dimensional array of
finite-state automata. The automata takes two states, 0 and 1, and
update their states simultaneously in discrete time by the same
cell-state transition function $f: \{0, 1\}^3 \rightarrow \{0,
1\}$. Each automaton updates its state depending on its current state
and states of its two closest neighbours.  When referring to
cell-state transition rules we use a decimal representation of the
cell-state transition table~\cite{wolfram_1983}; See examples in
\cite{wolfram} and extensive analysis of ECA's rules, parameters and
global transition graphs in~\cite{wuensche_1992}. Due to symmetries
the elementary transition rules can be grouped in the 88 classes with
equivalent behaviour~\cite{walker_1971,wuensche_1992}. We analyse, and
illustrate our discussions with, minimal decimal value rules from each
equivalence class. The ECA rules are studied using two statistical
measures: the Derrida coefficient and generative morphological
diversity.
 
The Derrida plot ~\cite{derrida_1986} is used in the evaluation of
Boolean networks~\cite{wuensche_1992, kauffman_1993, harris_2002,
  wuensche_2011}. The Derrida plot provides a statistical measure of
the divergence/convergence of network dynamics in terms of ÒHamming
distanceÓ $H$. The distance $H$ between two binary states of equal
size, $n$, is the number of sites that differ. The normalised Hamming
distance is $H/n$. The Derrida plot is calculated as
follows~\cite{wuensche_2011}. We randomly select a pair of initial
states, $c_1^0$ and $c_2^0$, separated by a small Hamming distance of
$H_0$ at time step $t=0$. We iterate the configurations using the same
cell-state transition rule for $m$ steps and measure $H$ between
configurations $c_1^m$ and $c_2^m$, repeat the measurement for more
samples pairs of initial configurations with the same $H_0$, and then
plot normalised $H_0$ against the mean normalised value of $H$. The
procedure is repeated for larger values of $H_0$.

The Derrida coefficient~\cite{wuensche_2011,ddlab_site}, analogous to
the Lyapunov exponent but for discrete systems, measures
sensitivity to initial conditions.  The Derrida coefficient is derived
from the initial slope $x$ of the Derrida plot.  For these results
$m$=1, initial $H_0$=1, increasing by 1 for 10 samples of 3000.  The
Derrida coefficient is calculated as $D=\log_2(\tan(x))$.  Boolean
networks and cellular automata behaving ``chaotically'' have positive
$D$, ordered dynamics have negative $D$. For Boolean networks, $D=0$
is attributed to dynamics at the edge of order and
chaos~\cite{harris_2002}, whereas for cellular automata $D=0$ merely
indicates stability.

Generative morphological diversity $\mu$ of a ECA characterises how
many different triplets, taken at time steps $t-1$, $t$ and $t+1$, of
neighbourhood configurations are generated by the ECA starting from a
single central cell in a state
1~\cite{adamatzky_martinez_2010,redeker_2013}. The measure is very
close to the in-degree histogram proposed in \cite{wuensche_1999}. We
have chosen $3\times3$ cell blocks to characterise morphology of
space-time configuration because a minimal block must include a cell
neighbourhood (three cells), include at least two subsequent local
configurations, to characterise identifiability, and sides
corresponding to time and space have the same number of cells. We
calculate morphological diversity $\mu$ using blocks of neighbourhood
states taken at three subsequent time steps: the automaton evolves for
$m$ steps list $\mathbf{L}$ of different $3\times3$ blocks from its
space-time configuration $c \times T$ is filled; $m$ is chosen
experimentally such that $L^m=L^{m-1}$.  The diversity
$\mu=|\mathbf{L}|$ is a size of list $\mathbf{L}$.

Values of $\mu$ and $D$ for representative rules of equivalence
classes are shown in Appendix, Tab.~\ref{data}.

\section{Creativity of ECA rules}

\begin{sidewaysfigure}[!tbp]
\includegraphics[width=1\textwidth]{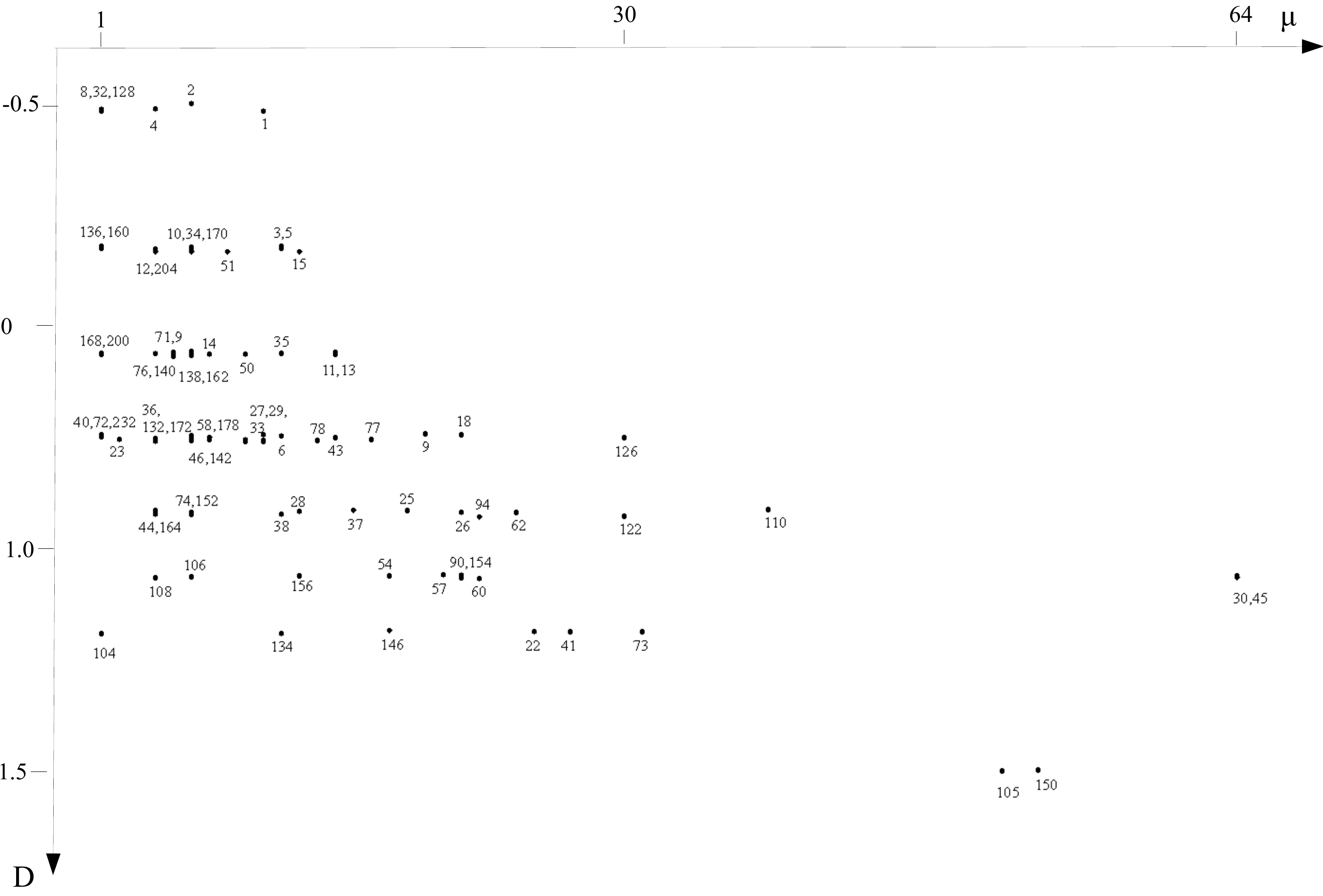}
\caption{Representative rules of 88 classes are displayed on
generative diversity $\mu$ versus Derrida coefficient $D$ plane.}
\label{functionnumbers}
\end{sidewaysfigure}

 \begin{sidewaysfigure}[!tbp]
\includegraphics[width=1\textwidth]{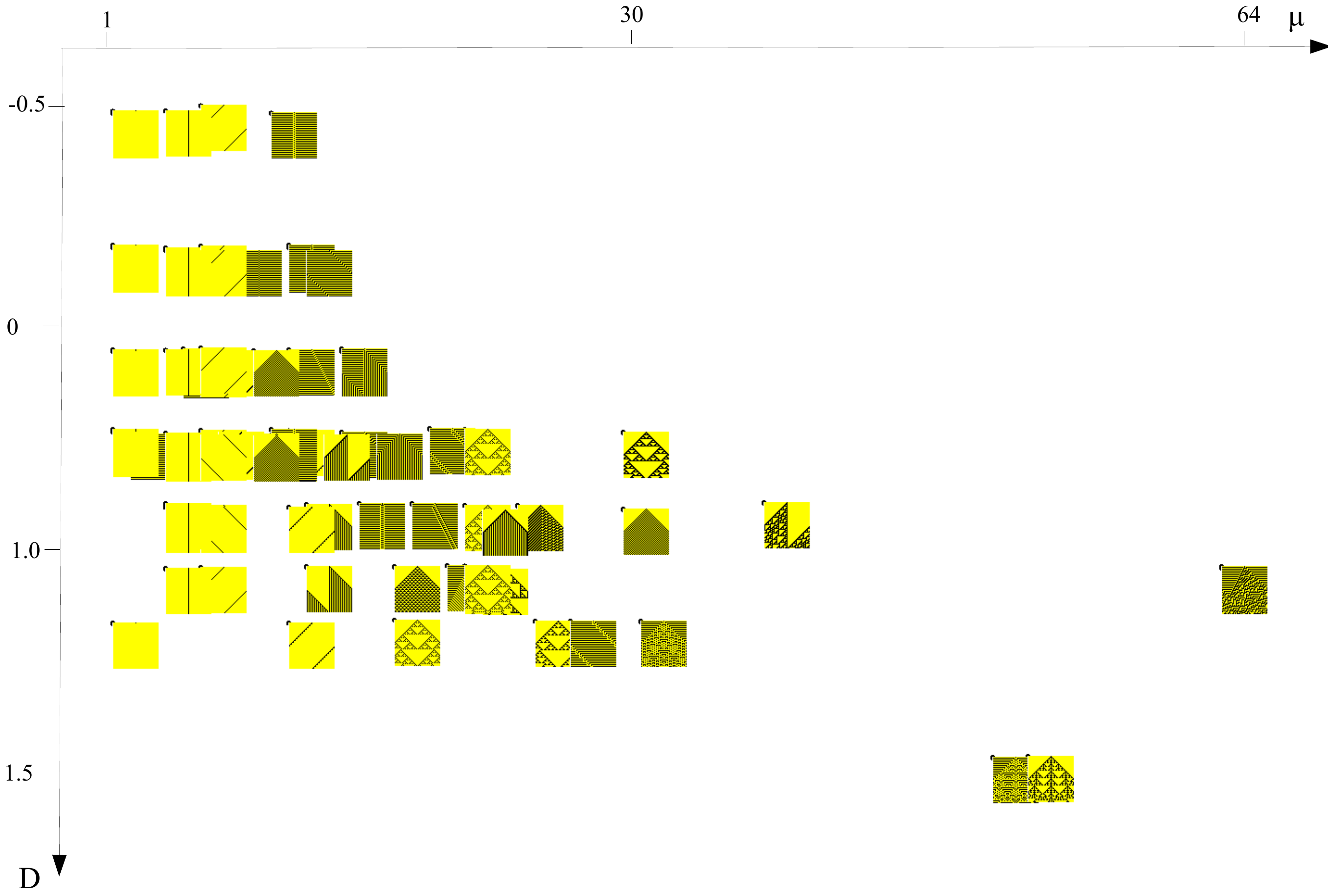}
\caption{Space-time configurations of representative rules of 88
  classes are displayed on generative diversity $\mu$ versus Derrida
  coefficient $D$ plane. Each automaton, 150 cells, started its
  development in configuration where all cells but one are in state 0
  and evolved for 150 iterations. Boundaries are periodic. Cells in
  state 1 are shown by black pixels, cells in state 0 are grey. }
\label{singlestartconfs}
\end{sidewaysfigure}

 \begin{figure}[!tbp]
\includegraphics[width=0.9\textwidth]{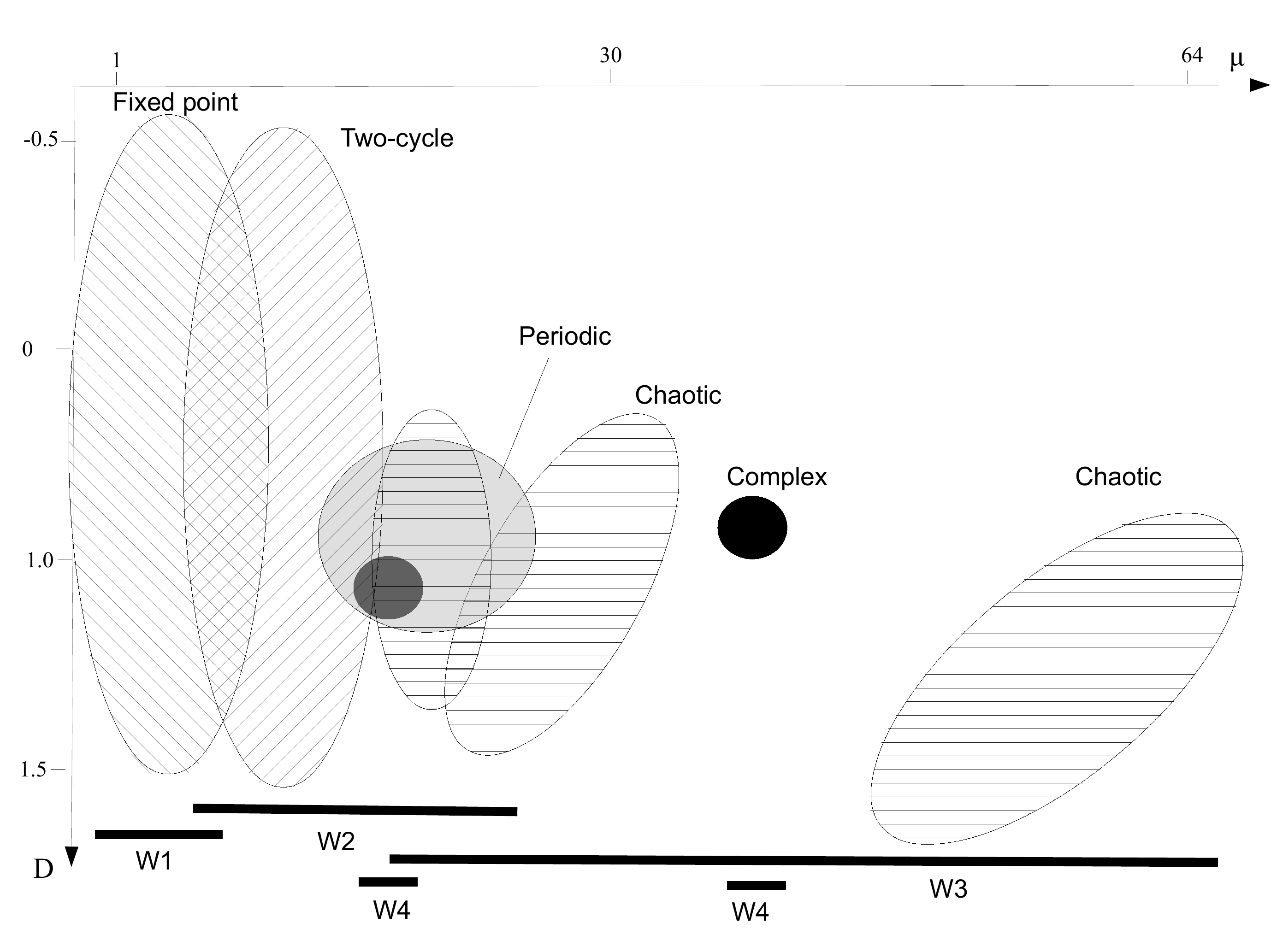}
\caption{Domains of main behavioural classes~\cite{oliveira_2001} in
  $\mu$--$D$ space. Projections of domains onto Wolfram
  classes~\cite{wolfram_1984a, wolfram_1984b} $W1$ to $W4$ are shown
  as solid thick lines. }
\label{clusters}
\end{figure}

Representative rules of the 88 equivalence classes are mapped onto
$\mu$-$D$ space in Fig.~\ref{functionnumbers}. Space-time
configurations, starting in configuration $0 \cdots 010 \cdots 0$,
generated by the rules from Fig.~\ref{functionnumbers} are shown in
Fig.~\ref{singlestartconfs}.  A substantial number of rules occupy a
domain of low values of $\mu$ yet spread more or less equally along
$D$ axis. Rules showing moderate generative diversity ($\mu$=20 to 40)
have Derrida coefficients around $D=1$.  Rules with highest generative
diversity ($\mu$=50 to 64) have values of $D$ ranging from nearly 1 to
1.6 (Fig.~\ref{functionnumbers}).  The increase in generative
diversity is visualised in sample configurations of representative
rules (Fig.~\ref{singlestartconfs}).

Domains of ECA behavioural classes~\cite{oliveira_2001} are shown in
Fig.~\ref{clusters}. Fixed point and two-cycle classes
~\cite{wolfram_1984a, wolfram_1984b} lie in the region of low
generative diversity yet fully spread along the Derrida coefficient
axis. Rules with periodic behaviour occupy a part of $\mu$--$D$ space
for average values of generative diversity and Derrida coefficient
equal to 1. Chaotic rules are spread from moderate to maximum values
of diversity and Derrida coefficient from 0.5 to 1.5.  Two complex
rules reside in a region of $\mu$ equals 1 and moderate and slightly
above average diversity $\mu$ (Fig.~\ref{clusters}).

Wolfram classes~\cite{wolfram_1984a, wolfram_1984b} $W1$ (fixed
point), $W2$ (periodic), $W3$ (chaotic) and $W4$ (complex) are well
arranged along the generative diversity axis, apart of class $W4$. One
rule of class $W4$ lies in the middle of class $W3$ and another rule
of class $W4$ lies in the intersection of classes $W2$ and $W3$
(Fig.~\ref{clusters}).

  \begin{figure}[!tbp]
\centering
\includegraphics[width=1.\textwidth]{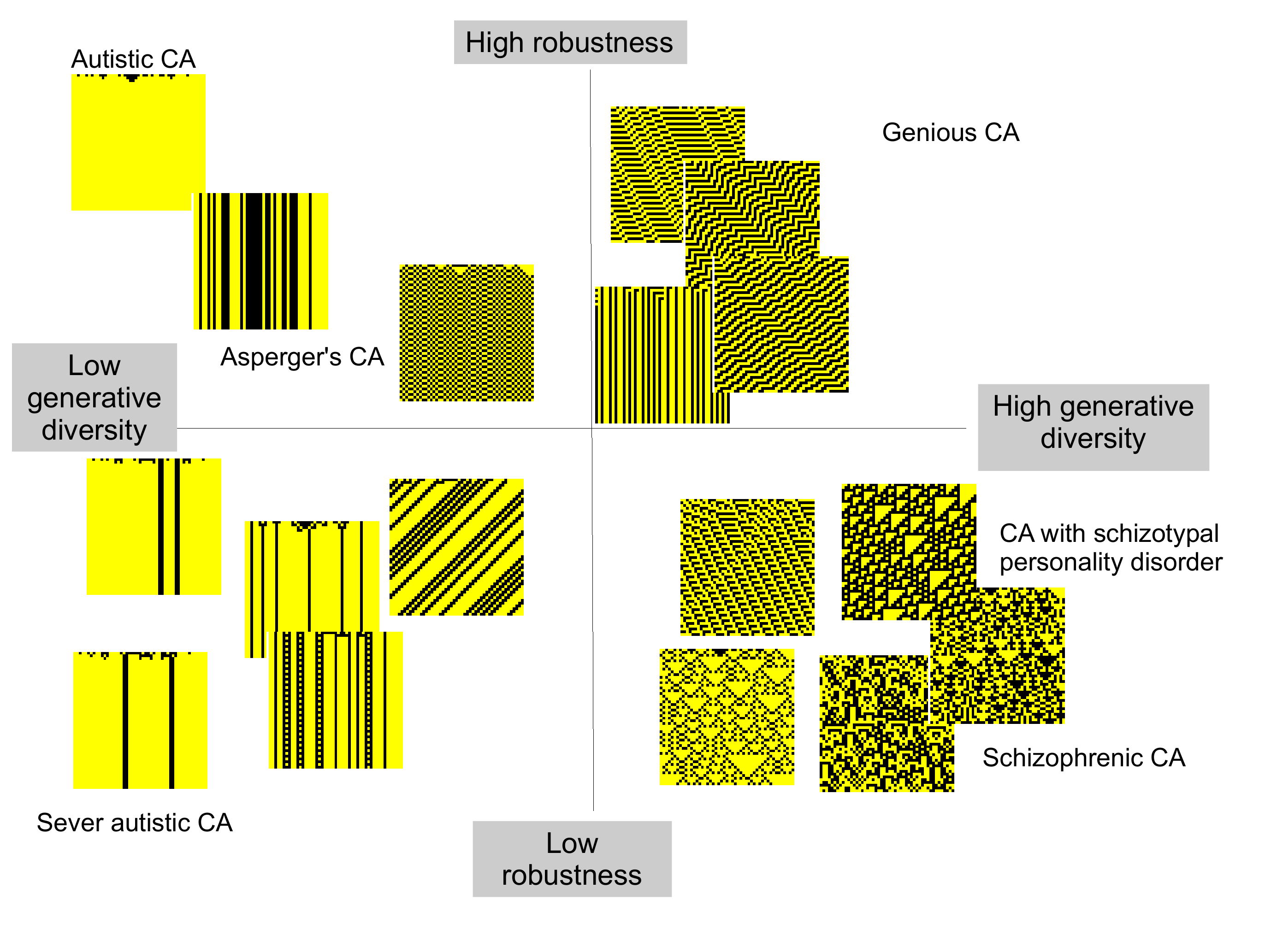}
\caption{Schyzotipy versus cognitive control spaces as seen via
  generative morphological diversity and robustness (Derrida
  coefficients). Interpretation of scheme Fig.~\ref{scheme} in terms
  of ECA. Examples of space-time configurations generated by autistic,
  creative and schizophrenic ECA rules. Configurations evolved from
  initially random uniform distribution of states 0 and 1. Cells in
  state 1 are black pixels, in state 0 are yellow/grey pixels. 
 }
\label{exampleconfi}
\end{figure}

From the distribution of rules (Fig.~\ref{functionnumbers}) and
domains of behavioural classes (Fig.~\ref{clusters}) we can speculate
that --- overall --- the increase in behavioural complexity, as
measured by generative diversity, leads to a decrease in robustness
and an increase in sensitivity to initial conditions, as measure by
the Derrida coefficient.

Ideally, highly articulated creative rules would appear in the upper right 
corner of the
upper right quadrant of the $\mu$--$D$ plane, but because this corner
is almost empty, we settled on rules closest to it. Such rules should
have above average generative morphological diversity, and below
average Derrida coefficients: $\mu > 11$ and $D < 0.53$ (we omit
rule 0 from calculating averages as not posing any interest). The
following equivalence classes, labelled by their representative rules,
satisfy the creativity condition: 3, 5, 11, 13, 15 and 35.
Equivalence classes 3 and 5 show the highest degree of robustness,
which represent cognitive control, amongst the creative rules with a
yet lower degree of generative diversity, representing the degree of
schizotypy. Equivalence classes 11 and 13 show higher generative
diversity yet lower robustness. Exemplar configurations of creative
ECA rules are shown in Fig.~\ref{exampleconfi}.  The creative ECA are
characterised by propagating patterns, which strikingly resemble waves
of excitation propagating in non-linear active media. There are
physiological correlations, see review in~\cite{kuszewski_2009}, that
creative individuals show activity in both hemispheres and increased
inter-hemispheric transfer.
 
In the quadrant of low generative diversity and high robustness
we observe a transition from normal ECA to Asperger's syndrome ECA to
autistic ECA (Fig.~\ref{exampleconfi}).  Normal rules, i.e. those
with $\mu$ and $D$ values closest to average, show stationary or
breathing domains of intermittent coherent patterns. Rules analogous
to Asperger's syndrome show configurations densely populated with
uniform, solid, domains of cells in 1 or 0.  ECA interpreted as
autistic evolve to fixed all-1 or all-0 global states.

Chaotic rules populate the quadrant corresponding to
schizophrenia and schizotypal personality disorders
(Fig.~\ref{exampleconfi}).  The most morphologically diverse and less
robust, and thus most 'schizophrenic', equivalence classes are 30, 45,
105 and 150. Rule 30 is a 'typical' chaotic rule, even used in a
random number generator~\cite{wolfram_2002}; when enriched with memory
rule 30 shows pronounced dynamics of gliders with sophisticated
interaction patterns~\cite{martinez_2010}.

Autistic ECA show stationary domains of alike states. There are no
propagating patterns in autistic ECA.  The stationary non-interacting
domains imitate zones of persistent nervous activity in a brain of a
severely autistic person. This could be a possible sign of
desynchronisation in motor cortex~\cite{welsh_2005, martineau_2008,
  rippon_2007}
  
The dynamics of ECA governed by schizophrenic rules is characterised
by sudden emergence and subsequent swift collapse of domains of alike
states. These are reflected in triangular tessellations visible in
space-time configurations (Fig.~\ref{exampleconfi}). Assume that a
one-dimensional ECA is an abstraction of a brain, and that patterns of
1s are analogous of neurons bursting with excitation spikes. Then a
creative brain produces coherent yet morphologically rich pattens of
nervous activity, e.g. propagating auto-waves, while a brain with high
schizophrenic disorder shows (quasi-) chaotic, incoherent and
`spontaneous' outburst of nervous activity. These outburst of activity
imitate abnormalities in multiple parts of the brain and diminished
temporal stability~\cite{kasai_2002,basile_2004, jordanov_2011}.

 \section{Discussion}

  \begin{table}
  \centering
  \caption{Four classes of CA creativity.}
  \begin{tabular}{l|l}
  Class & Rules \\ \hline
  Creative & 
 3, 
5, 
11, 
13, 
15, 
35 \\
Schizophrenic & 
9, 
18, 
22, 
25, 
26, 
28, 
30, 
37, 
41, 
43, 
45, 
54, 
57, 
60, 
62, 
73, 
77, 
78, 
90, 
94,  \\
 & 
105, 
110, 
122, 
126, 
146, 
150, 
154, 
156\\
Autistic savants &
1, 
2, 
4, 
7, 
8, 
10, 
12, 
14, 
19, 
32, 
34, 
42, 
50, 
51, 
76, 
128, 
136, 
138, 
140, 
160, \\
& 
162, 
168, 
170, 
200, 
204\\
Severely autistic &
23, 
24, 
27, 
29, 
33, 
36, 
40, 
44, 
46, 
56, 
58, 
72, 
74, 
104, 
106, 
108, 
130, 
132, \\
& 
142, 
152, 
164, 
172, 
178, 
184, 
232
  \end{tabular}
  \label{tablecreativeclasses}
  \end{table}

Using measures of generative morphological diversity and the Derrida
coefficient we classified ECA rules onto a spectrum of autistic,
schizophrenic and creative personality.  Four classes are shown in 
Tab.~\ref{tablecreativeclasses}.

Autistic rules correspond to rule classes with fixed point behaviour, schizophrenic rules are
chaotic and creative rules belong to a class of two-cycle
behaviour. There are two types of creativity: creative product and
creative process~\cite{maher_2012}. The creative ECA rules discovered
correspond to a creative process; space-time configurations produced
by a creative rule may not be creative.  Rule 54 and 110 are
computationally universal~\cite{wolfram_2002, cook_2004,
  martinez_2006, martinez_2011a} but why are they not creative?
Because they lack robustness, autonomous cognitive control. These
rules perform computation only with strict initial conditions. The
computational circuits in these rules do not emerge in their
space-time configurations by themselves.

We are aware that this interpretation will appear too simplistic, and
that both personality and cellular automata are profoundly
complex. However, we decided to develop this naive conceptual
approach to provoke new ways of thinking and discussion about the issues. We also, believe that
highly articulated creative rules might be found in a richer rule-space than ECA.

\newpage

\section*{Appendix}
\begin{table}[!h]
\caption{Values of $\mu$ and $D$ for representative rules of equivalence classes.}
\centering
{\scriptsize
\begin{tabular}{ll}
\begin{tabular}{c|cc}
Rule	&	$\mu$ 	&	$D$	\\	\hline										
0	&	1	&	-inf	\\												
1	&	10	&	-0.423	\\												
2	&	6	&	-0.446	\\												
3	&	11	&	-0.017	\\												
4	&	4	&	-0.431	\\												
5	&	11	&	-0.009	\\												
6	&	11	&	0.557	\\												
7	&	5	&	0.303	\\												
8	&	1	&	-0.424	\\												
9	&	19	&	0.55	\\												
10	&	6	&	-0.007	\\												
11	&	14	&	0.304	\\												
12	&	4	&	-0.007	\\												
13	&	14	&	0.311	\\												
14	&	7	&	0.309	\\												
15	&	12	&	0	\\												
18	&	21	&	0.553	\\												
19	&	5	&	0.317	\\												
22	&	25	&	1.143	\\												
23	&	2	&	0.566	\\												
24	&	6	&	0.567	\\												
25	&	18	&	0.782	\\												
26	&	21	&	0.786	\\												
27	&	10	&	0.573	\\												
28	&	12	&	0.783	\\												
29	&	10	&	0.569	\\												
30	&	64	&	0.98	\\												
32	&	1	&	-0.424	\\												
33	&	10	&	0.552	\\												
34	&	6	&	-0.013	\\												
35	&	11	&	0.307	\\												
36	&	4	&	0.564	\\												
37	&	15	&	0.78	\\												
38	&	11	&	0.792	\\												
40	&	1	&	0.553	\\												
41	&	27	&	1.145	\\												
42	&	6	&	0.313	\\												
43	&	14	&	0.561	\\												
44	&	4	&	0.792	\\												
45	&	64	&	0.976	\\												
46	&	7	&	0.567	\\												
50	&	9	&	0.309	\\												
51	&	8	&	0	\\												
54	&	17	&	0.975	\\												
56	&	6	&	0.786	\\												
57	&	20	&	0.972	\\
\end{tabular}
&
\begin{tabular}{c|cc}
Rule	&	$\mu$ 	&	$D$ 	\\ \hline											
58	&	9	&	0.568	\\												
60	&	22	&	0.983	\\												
62	&	24	&	0.787	\\												
72	&	1	&	0.557	\\												
73	&	31	&	1.144	\\												
74	&	6	&	0.792	\\												
76	&	4	&	0.308	\\												
77	&	16	&	0.567	\\												
78	&	13	&	0.57	\\												
90	&	21	&	0.982	\\												
94	&	22	&	0.8	\\												
104	&	1	&	1.15	\\												
105	&	51	&	1.564	\\												
106	&	6	&	0.979	\\												
108	&	4	&	0.981	\\												
110	&	38	&	0.778	\\												
122	&	30	&	0.799	\\												
126	&	30	&	0.561	\\												
128	&	1	&	-0.429	\\												
130	&	6	&	0.555	\\												
132	&	4	&	0.563	\\												
134	&	11	&	1.149	\\												
136	&	1	&	-0.009	\\												
138	&	6	&	0.312	\\												
140	&	4	&	0.308	\\												
142	&	7	&	0.56	\\												
146	&	17	&	1.14	\\												
150	&	53	&	1.562	\\												
152	&	6	&	0.793	\\												
154	&	21	&	0.974	\\												
156	&	12	&	0.975	\\												
160	&	1	&	-0.017	\\												
162	&	6	&	0.302	\\												
164	&	4	&	0.782	\\												
168	&	1	&	0.307	\\												
170	&	6	&	0	\\												
172	&	4	&	0.572	\\												
178	&	9	&	0.574	\\												
184	&	6	&	0.571	\\												
200	&	1	&	0.311	\\												
204	&	4	&	0	\\												
232	&	1	&	0.558	\\
\end{tabular}\\
\end{tabular}
}
\label{data}
\end{table}

\end{document}